\documentstyle[aps,psfig]{revtex}
\tightenlines

\begin{document}
\title{Mechanical response functions of finite temperature Bose-Einstein Condensates}
\author{Stephen Choi$^{1}$, Vladimir Chernyak$^{2}$, and Shaul Mukamel$^{1,3}$}
\address{$^{1}$ Department of Chemistry, University of Rochester, Box 270216, Rochester, New York 14627-0216 \\
$^{2}$ Corning Incorporated, Process Engineering and Modeling, Corning, New York 14831 \\
$^{3}$ Department of Physics and Astronomy, University of Rochester, Box 270216, Rochester, New York 14627-0216}

\maketitle

\begin{abstract}
Using the Liouville space framework developed in nonlinear optics we calculate the linear response functions and susceptibilities of Bose-Einstein condensates (BEC)  subject to an arbitrary  mechanical force. 
Distinct signatures of the dynamics of finite temperature BEC  are obtained by solving the Hartree-Fock-Bogoliubov theory. Numerical simulations of the position dependent linear response functions of one dimensional trapped BEC in the time and the frequency domains are presented. %\\

%\vspace{0.9cm} \\
%{\it Submitted to Phys. Rev. A}
\end{abstract}


\vspace{6mm}

\section{Introduction}

Since the first experimental observation of atomic Bose-Einstein Condensates (BEC), extensive theoretical and experimental effort has focused on understanding their properties\cite{history}. BEC presents many possible  avenues of research, as it may be viewed in a variety of ways such as  mesoscopic ``superatom'',  gaseous ``superfluid'', or as a source of coherent atoms used in an Atom Laser. In particular, the close analogy between the nonlinear interaction of BEC matter-waves and photon waves in nonlinear optics gave rise to fascinating atom optics applications\cite{meystre}.

Optical spectroscopy has been a major tool for studying the properties of matter since the early days of quatum physics. And, with the advent of the laser, nonlinear optical spectroscopy has become an important technique for studying the properties of matter that are not accessible with  incoherent light sources.  The primary theoretical tool used in nonlinear optical spectroscopy to analyze the structure and dynamic processes in many body quantum systems is the linear and higher order optical response functions\cite{spectroscopy}.  
In this paper we extend the systematic formalism of optical response functions to probe the response of trapped, atomic BEC  to an arbitrary {\it mechanical force} coupled to the atomic density.  We calculate the  first order (linear) suseptibilities for the condensate, non-condensate density, and non-condensate correlation in the time and the frequency domains.

A major difference that arises  when the formalism of nonlinear spectroscopy is applied to the mesoscopic BEC is that it is generally not possible to apply the dipole approximation commonly made in the atom-light interaction.  It should also be noted that there have been a number of calculations of the linear response of BEC in the past\cite{Giorgini}; the {\em response functions} calculated here provide a more fundamental Green's function description of the response to a force with arbitrary spatial and temporal profiles. The time domain response functions or their frequency domain counterparts, the susceptibilities, provide unique signatures of the dynamics of the system under consideration.

The dynamics of zero temperature atomic BEC is commonly described using the time-dependent and time-independent Gross-Pitaevskii Equation (GPE). Numerical solutions of the GPE for various properties of zero temperature BEC as well as for  atom optics and four wave mixing applications have been reported\cite{Dalfovo,Edwards1,atomoptics,Band}.  BEC at finite temperatures, on the other hand, require sophisticated theories that go beyond the GPE, and various approaches including the time-dependent Bogoliubov-de Gennes equations\cite{CastinDum}, the Hartree-Fock-Bogoliubov (HFB) theory\cite{Ring,Blaizot_Ripka,Griffin,Prou2}, Quantum Kinetic Theory\cite{Baym,Martin,Gardiner,ZNG,Walser,Zaremba}, and Stochastic methods\cite{Krauth,Ceperley,Walls,Drummond,Carusotto} have been employed.

In order to go beyond the GPE, two or more particle correlations must be taken into account. One of the original contributions was made by Bogoliubov with his introduction of the Bogoliubov transformation\cite{Huang,Noziers} which shows how the condensed state of an interacting homogeneous gas differs from that of the noninteracting gas. That result was extended to inhomogeneous gases by de Gennes\cite{Noziers}. Recently, a time-dependent version of the Bogoliubov-de Gennes equations was employed by Castin and Dum\cite{CastinDum} to describe the dynamics of BEC in time dependent traps. Their approach was based on an expansion of the evolution equations for the atomic field operator which is valid if the number of noncondensate particles is small. Their result has been used for analyzing the stability and depletion of a strongly driven BEC\cite{CastinDum}.

The time-independent HFB theory (TIHFB) has been used to calculate some of the important equilibrium properties such as the quasiparticle excitation frequencies and the equilibrium condensate and non-condensate density profiles\cite{Griffin}. The dynamical properties of finite temperature BEC predicted by the TDHFB\cite{Prou2} have not been studied extensively, because the full solution to these equations is computationally prohibitive. The response functions computed here offer a systematic perturbative approach for exploring the TDHFB dynamics, at an affordable numerical cost.

The quantum kinetic theory of dilute interacting Bose gas was derived long ago by Kadanoff and Baym\cite{Baym} in terms of nonequilibrium real-time Green's functions that parametrize the condensating gas. Their equations were later generalized by Hohenberg and Martin to include condensates\cite{Martin}.  A more contemporary version of the quantum kinetic theory applicable to the experimentally produced BEC has been developed in a series of papers of Gardiner and Zoller\cite{Gardiner}. They describe a system composed of interacting a condensate and a noncondensate vapor, where the vapor is described by a quantum kinetic master equation, equivalent to a quantum Boltzmann equation of the Uehling-Uhlenbeck form\cite{Uhlenbeck}. The master equation which describes the transfer of energy and particles between the vapor and condensate has been used to describe the formation of BEC.  A similar formalism was also put forward by Walser {\it et al.}\cite{Walser} who derived, using a method reminiscent of the classical Bogoliubov-Born-Green-Kirkwood-Yvon (BBYGK) technique, a generalized kinetic theory for the coarse-grained Markovian many-body density operator. Their result incorporates second order collisional processes, and describes irreversible evolution of a condensed bosonic gas of atoms towards thermal equilibrium.

Important experimental observations such as the damping of elementary excitations has prompted attempts to develop a theory that works in the hydrodynamic regime. These have led to various versions of Quantum Kinetic theory such as the two fluid hydrodynamic description of finite temperature BEC\cite{ZNG} developed by Zaremba, Nikuni, and Griffin (ZNG). ZNG is a semiclassical Hartree Fock description where one neglects the off-diagonal distribution functions. It combines a non-Hermitian generalized Gross-Pitaevskii Equation for the condensate wave function and the Boltzmann kinetic equation for the noncondensate phase density. This theory was recently shown to properly describe damping\cite{Zaremba}.

It will be helpful to clarify how the various kinetic theories are related. The Green's function approach of Kadanoff and Baym\cite{Baym} are equivalent to the equations of Walser {\it et al.}\cite{Walser}.  From the equations of Walser {\it et al.}, the ZNG hydrodynamic equations may be obtained by neglecting the anomalous fluctuations. This greatly simplifies the equations and facilitates the numerical solution. The TDHFB equations are obtained by dropping the second order collisional terms from the kinetic equations of Walser {\it et al.} while keeping the anomalous fluctuations.  The theory of Gardiner and Zoller\cite{Gardiner} is based on the quantum Boltzmann master equations reminiscent of the quantum stochastic methods used in Quantum Optics; their quantum kinetic master equations also contain the higher order collisional processes described by Walser {\it et al.} and ZNG. A number of approximations are common to these kinetic theories: the Born and Markov approximation, ergodic assumption, and Gaussian initial reference distribution. The validity of these assumptions will ultimately be confirmed by experiments; so far the experiments have supported the predictions of these equations.

The stochastic method is another approach used for the description of finite temperature BEC beyond GPE. A description of a Bose gas in thermal equilibrium has been developed using the quantum Monte Carlo techniques, based on Feynman path integral formulation of quantum mechanics\cite{Krauth,Ceperley}. A stochastic scheme corresponding to the dynamical evolution of density operator in the positive $P$ representation has been applied to BEC\cite{Walls,Drummond}. An alternative approach that describes the dynamics of the gas based on a stochastic evolution of Hartree states and avoids some of the instability problems of earlier works was proposed recently\cite{Carusotto}. 
The stochastic methods make the simulation of the exact dynamics of $N$ boson system numerically feasible; this, at present, requires assuming an initial state such as Hartree Fock state which is not very realistic.

The TDHFB theory is a self-consistent theory of BEC in the collisionless regime that progresses logically from the Gross-Pitaevskii Equation by taking into account higher order correlations of noncondensate operators. Although TDHFB does not take into account higher order correlations that are done in the various quantum kinetic theories, the TDHFB equations are valid at temperatures near zero, even down to the zero temperature limit, and are far simpler than the kinetic equations which can only  be solved using approximations such as ZNG. Another attractive feature of TDHFB from a  purely pragmatic point of view is that the Fermionic version of the theory has already been well-developed in Nuclear Physics\cite{Ring}. We therefore work at the TDHFB level  in this paper and our approach draws upon the analogy with the time-dependent Hartree-Fock (TDHF) formalism developed for nonlinear optical response of many electron systems\cite{TDHF}.

The paper is organized as follows: in Section II we show how to systematically solve the TDHFB equations for externally driven finite temperature BEC by order by order expansion of the dynamical variables in the external field. In Section III we define the $n$'th order response function in time and frequency domains and calculate the linear ($n = 1$) response function.
Numerical results  for the linear response functions and susceptibilities of a 2000 atom condensate in a one dimensional harmonic trap, and its variation with position, time and frequency, are presented in Section IV at zero and finite temperatures.  Our main findings are finally summarized in Section V.

\section{The TDHFB Equations}

\subsection{Equations of motion}
Our theory starts with a time-dependent many-body second quantized Hamiltonian describing a  system of externally driven, trapped,
structureless bosons with pairwise interactions. Introducing the boson operators $\hat{a}^{\dagger}_{i}$
and $\hat{a}_{i}$ that respectively create and annihilate a particle from
a basis state $i$ with wave functions $\phi_{i}({\bf r})$, the
Hamiltonian is:
\begin{equation}
\hat{H} = \hat{H}_0 +  \hat{H}^{\prime}(t)
\end{equation}
where
\begin{equation}
\hat{H}_0 = \sum_{ij} \left ( H_{ij} - \mu \right ) \hat{a}^{\dagger}_{i}\hat{a}_{j} + \frac{1}{2}\sum_{ijkm} V_{ijkl} \hat{a}^{\dagger}_{i}\hat{a%
}^{\dagger}_{j}\hat{a}_{k}\hat{a}_{l}.  \label{Hamiltonian}
\end{equation}
%\begin{equation}
%H^{\prime\prime} =  \sum_{i} \lambda^{*}_{i}\hat{a}_{i} +  %\lambda_{i}\hat{a}^{\dagger}_{i}.
%\end{equation}
The matrix elements of the single particle Hamiltonian $H_{ij}$ are given by
\begin{equation}
H_{ij} = \int \! d^{3}{\bf r} \, \phi^{*}_{i}({\bf r}) \left [ - \frac{\hbar^2}{2m}
\nabla^{2} + V_{\mbox{\scriptsize trap}}({\bf r}) \right ]  \phi_{j}({\bf r}),
\end{equation}
where $V_{\mbox{\scriptsize trap}}({\bf r})$ is the magnetic potential that confines
the atoms and $\mu$ is the chemical potential. The basis state $\phi_{i}({\bf r})$ is arbitrary; a convenient basis for trapped BEC is the eigenstates of the trap since $H_{ij}$ is then diagonal.  The symmetrized two particle interaction matrix elements in Eq.~(\ref
{Hamiltonian}) are
\begin{equation}
V_{ijkl} = \frac{1}{2} \Big [ \mbox{$\langle i
j |$}V\mbox{$| k l \rangle$} + \mbox{$\langle j i |$}V\mbox{$| k l \rangle$} %
\Big ],
\end{equation}
where
\begin{equation}
\begin{array}{c@{\hspace{1cm}}c}
{\displaystyle \mbox{$\langle i j |$}V\mbox{$| k l \rangle$} = \int \! d^{3}%
{\bf r} \, d^{3}{\bf r^{\prime}} \, \phi^{*}_{i}({\bf r})\phi^{*}_{j}({\bf %
r^{\prime}})V({\bf r}-{\bf r^{\prime}})\phi_{k}({\bf r^{\prime}})\phi_{l}(%
{\bf r}) , } & \label{Vijkl}
\end{array}
\end{equation}
with $V({\bf r}-{\bf r^{\prime}})$ being a general interatomic potential.
$H^{\prime}(t)$ describes the coupling of an external field with the system:
\begin{equation}
H^{\prime}(t) = \eta \sum_{ij} E_{ij}(t)\hat{a}^{\dagger}_{i}\hat{a}_{j}. \label{ext_perturb}
\end{equation}
$\eta$ is a  bookkeeping expansion parameter (to be set to 1 at the end of the calculation). 
The matrix elements $E_{ij}(t)$ are given by 
\begin{equation}
E_{ij} = \int \! d^{3}{\bf r} \, \phi^{*}_{i}({\bf r}) V_{f}({\bf r},t)
\phi_{j}({\bf r}), \label{Eij}
\end{equation}
where $V_{f}({\bf r},t)$ denotes a time- and position- dependent external potential that exerts an arbitrary mechanical force on the system.

More general forms of $H^{\prime}(t)$ could include, for example, terms of the form $\sum_{ij} F_{ij}(t)\hat{a}_{i}\hat{a}_{j}$ in addition to the term given in Eq. (\ref{ext_perturb}) which couples to atomic density. In this work we shall focus on the most experimentally relevant perturbation; for instance, the time-dependent modulation of the trap spring constant which mimics the mechanical force that was recently applied
experimentally\cite{ExRb,ExNa}.

The dynamics of the system is calculated by deriving equations of motion for the condensate mean field, $z_{i} \equiv  \langle \hat{a}_{i} \rangle$, the non-condensate density $\rho_{ij} \equiv \langle \hat{a}_{i}^{\dagger} \hat{a}_{j} \rangle - \langle \hat{a}_{i}^{\dagger} \rangle \langle \hat{a}_{j} \rangle$, and the non-condensate correlations $\kappa_{ij} \equiv  \langle \hat{a}_{i} \hat{a}_{j} \rangle - \langle \hat{a}_{i} \rangle \langle \hat{a}_{j} \rangle$. 
Closed equations are derived starting with the Heisenberg equations of motion for the operators,
$\hat{a}_{i}$, $\hat{a}_{i}^{\dagger} \hat{a}_{j} $, and $\hat{a}_{j} \hat{a}_{j}$, 
 and assuming a coherent many body state. The resulting many body hierarchy is then truncated using the generalized Wick's theorem for ensemble averages\cite{WICK,choi2,Popov}:
\begin{eqnarray}
\langle A_{i} \rangle & \neq & 0,  \label{Wick1} \\
\langle A_{1} A_{2} \rangle &  = & \langle A_{1} \rangle \langle A_{2} \rangle + \langle \langle A_{1} A_{2} \rangle \rangle   \label{Wick2} \\
\langle A_{1} A_{2} A_{3} \rangle &  = & \langle A_{1} \rangle \langle A_{2} \rangle   \langle A_{3} \rangle  + \langle A_{1} \rangle \langle \langle A_{2} A_{3} \rangle \rangle +
\langle A_{2} \rangle \langle \langle A_{1} A_{3} \rangle \rangle +
\langle A_{3} \rangle \langle \langle A_{1} A_{2} \rangle \rangle  , \label{Wick3}
\end{eqnarray}
and similarly for products involving higher number of operators.  $A_{i}$ denote boson creation or annihilation operators, $\hat{a}^{\dagger}_{i}$ or $\hat{a}_{i}$, and are assumed to be normally ordered. We follow the convention that normal ordered operators are time ordered. 
The double angular brackets $\langle \langle A_{i}A_{j} \rangle \rangle$ denote 
 irreducible two point correlations. Using this notation we have $\rho_{ij} \equiv \langle \langle \hat{a}_{i}^{\dagger} \hat{a}_{j}  \rangle  \rangle $ and $ \kappa_{ij} \equiv \langle \langle \hat{a}_{i} \hat{a}_{j}  \rangle  \rangle $.

This procedure yields the TDHFB equations of motion\cite{Ring,Blaizot_Ripka,Prou2}
\begin{eqnarray}
i \hbar \frac{d z}{dt} & = & \left [ {\cal H}_{z} + \eta E(t) \right ] z + {\cal %
H}_{z*} z^{*} ,   \label{zdot} \\
i \hbar \frac{d \rho}{dt} & = & [h, \rho] - (\kappa \Delta^{*} - \Delta
\kappa^{*}) + \eta [E(t), \rho ] ,  \label{rhodot} \\
i \hbar \frac{d \kappa}{dt} & = & (h \kappa + \kappa h^{*}) + (\rho \Delta +
\Delta \rho^{*}) + \Delta + \eta [E(t),\kappa]_{+} ,  \label{kappadot}
\end{eqnarray}
where $[ \dots ]_{+}$ denotes the anticommutator.
Here,  ${\cal H}_{z}$ , ${\cal H}_{z*}$, $h$, and $\Delta$ are $n \times n$ matrices  with $n$ being the basis set size used:
\begin{eqnarray}
\left [ {\cal H}_{z} \right ]_{i,j} & = & H_{ij} - \mu + \sum_{kl} V_{iklj} \left [
z^{*}_{k}z_{l} + 2 \rho_{lk} \right ], \label{h_z} \\
\left [ {\cal H}_{z*} \right ]_{i,j} & = & \sum_{kl} V_{ijkl} \kappa_{kl}, \label{h_z_star} \\
h_{ij} & = & H_{ij} - \mu + 2 \sum_{kl} V_{iklj}  \left [
z_{k}^{*}z_{l} + \rho_{lk} \right ], \label{h} \\
\Delta_{ij} & = & \sum_{kl}  V_{ijkl}  \left[ z_{k} z_{l}
+ \kappa_{kl} \right]~.  \label{Delta}
\end{eqnarray}
$h$ is known as the Hartree-Fock
Hamiltonian and $\Delta$ as the pairing field\cite{Blaizot_Ripka}. $\mu$ is the chemical potential introduced in Eq. (\ref{Hamiltonian}). Eqs. (\ref{rhodot}) and (\ref{kappadot}) may also be recast in a more compact matrix form\cite{Blaizot_Ripka}:
%\begin{equation}
%i \hbar \frac{d}{dt}  \left (
%\begin{array}{c}
%z \\
%z^{*}
%\end{array}
%\right ) = \left (
%\begin{array}{cc}
%{\cal H}_{z} + \eta E(t) &  {\cal H}_{z*} \\
%-{\cal H}_{z*}^{*} & -\left [ {\cal H}_{z} + \eta E(t) \right ]^{*}
%\end{array}
%\right ) \left (
%\begin{array}{c}
%z \\
%z^{*}
%\end{array}
%\right ) ,  \label{matrixnlse}
%\end{equation}
\begin{equation}
i \hbar \frac{d \tilde{\gamma}R}{dt} = \tilde{\gamma} [H \tilde{\gamma}, \tilde{\gamma} R] + \eta E , \label{TDHFB}
\end{equation}
%where Eq. (\ref{matrixnlse})  is the time-independent nonlinear
%Schr\"{o}dinger equation (NLSE) with ${\cal H}_{z}$ and ${\cal H}_{z*}$ being 
%the effective Hamiltonians defined by Eqs. (\ref{h_z}-\ref{h_z_star}). 
where $H$, $R$, $E$ and $\tilde{\gamma}$ are $2n \times 2n$ matrices defined as follows:
\begin{equation}
H = \left (
\begin{array}{cc}
h - \mu &  \Delta \\
\Delta^{*} & h^{*} - \mu
\end{array}
\right )  \;\;\;\;  R = \left (
\begin{array}{cc}
\rho & \kappa \\
\kappa^* & \rho^* + \openone
\end{array}
\right )  \;\;\;\;
E = \left (
\begin{array}{cc}
[E(t), \rho] & [E(t), \kappa]_{+} \\
-[E(t), \kappa]_{+}^* & -[E(t), \rho]^*
\end{array}
\right ),   \label{HR}
\end{equation}
\begin{equation}
\tilde{\gamma} = \left (
\begin{array}{cc}
\openone & 0 \\
0 & -\openone
\end{array}
\right ),   \;\;\;\;  \tilde{\gamma}^2 =  \tilde{\gamma},
\end{equation}

\subsection{Solutions of the TDHFB}

Direct numerical solution of the TDHFB equations [Eqs (\ref{zdot} - \ref{Delta})] for arbitrary field strength is complicated by their highly nonlinear character. A practical way to proceed is expanding all dynamical variables in powers of $\eta$:
\begin{equation}
z_{i}(t) = z^{(0)}_{i} + \eta z^{(1)}_{i}(t) + \eta^{2} z^{(2)}_{i}(t) +
\cdots   \label{zexp}
\end{equation}
\begin{equation}
\rho_{ij}(t) = \rho^{(0)}_{ij} + \eta \rho^{(1)}_{ij}(t) + \eta^{2}
\rho^{(2)}_{ij}(t) + \cdots \label{rhoexp}
\end{equation}
\begin{equation}
\kappa_{ij}(t) = \kappa^{(0)}_{ij} + \eta \kappa^{(1)}_{ij}(t) + \eta^{2}
\kappa^{(2)}_{ij}(t) + \cdots \label{kappaexp}
\end{equation}
and upon substituting these expansions in Eqs. (\ref{zdot} - \ref{Delta}), collecting all terms to a given power in $\eta$. This results in a hierarchy of equations such that the equation of motion for the $n$'th order solution $\alpha^{(n)}$, where $\alpha$ denotes the variables $z$, $\rho$ or $\kappa$, is expressed as a function of $0, \ldots (n-1)$th order solutions, $\alpha^{(0)}, \alpha^{(1)}, \ldots , \alpha^{(n-1)}$.  While the zero'th order equation is nonlinear, each of the higher order equations are linear; the $n$'th order solutions, $\alpha^{(n)}$ are therefore obtained by solving the sequence of
linear equations. This is analogous to the TDHF\cite{TDHF} or time dependent density functional algorithms used for many electron systems.

Finding the zero'th order solution $\{ z^{(0)}, \rho^{(0)}, \kappa^{(0)} \}$ should be the first step in solving the TDHFB.  This describes the system at equilibrium in the absence of the external driving field, and requires the solution of the TIHFB equations:
\begin{equation}
{\cal H}^{(0)}_{z} z ^{(0)} + {\cal H}^{(0)}_{z*} z^{(0)*}   =  0 \label{z0dot}
\end{equation}
\begin{equation}
[h^{(0)} , \rho^{(0)} ] - (\kappa^{(0)}  \Delta^{(0)*} - \Delta^{(0)} \kappa^{(0)*})  =  0 \label{rho0dot}
\end{equation}
\begin{equation}
(h^{(0)}  \kappa^{(0)}  + \kappa^{(0)}  h^{(0)*}) + (\rho^{(0)}  \Delta^{(0)}  +
\Delta^{(0)}  \rho^{(0)*}) + \Delta^{(0)}  =  0 \label{kappa0dot}
\end{equation}
Here,  ${\cal H}^{(0)} _{z}$ , ${\cal H}^{(0)} _{z*}$, $h^{(0)} $, and $\Delta^{(0)}$ are $n \times n$ matrices  defined in Eq. (\ref{h_z}-\ref{Delta}) with the variables $\{ z, \rho, \kappa \}$ replaced by $\{ z^{(0)}, \rho^{(0)}, \kappa^{(0)} \}$. It follows from Eq. (\ref{TDHFB}) that Eqs. (\ref{rho0dot}) and (\ref{kappa0dot}) can be written in the compact form:
\begin{equation}
\tilde{\gamma} [H^{(0)}\tilde{\gamma}, \tilde{\gamma} R^{(0)}] = 0. 
\end{equation}
Self-consistent numerical methods for solving the TIHFB are given in the Appendix\cite{Blaizot_Ripka}.

 So far we have worked in the trap basis since this enables us to maintain a general form for the interatomic interactions $V_{ijkl}$ and makes the numerical solution feasible. However, it may be more interesting to recast the solution in real space $\alpha^{(n)}({\bf r},t)$ ($\alpha =z$, $\rho$, or $\kappa$) by transforming the solution to TDHFB in the trap basis $\alpha^{(n)} (t)$: 
\begin{eqnarray}
z^{(n)} ({\bf r}, t) & = & \sum_{j} z^{(n)}_{j}(t) \phi_{j} ({\bf r}), \label{z_r} \\
\rho^{(n)} ({\bf r}, t) & = & \sum_{ij} \rho^{(n)}_{ij}(t) \phi^{*}_{i} ({\bf r})\phi_{j} ({\bf r}),   \label{rho_r} \\
\kappa^{(n)} ({\bf r}, t) & = & \sum_{ij} \kappa^{(n)}_{ij}(t) \phi_{i} ({\bf r}) \phi_{j} ({\bf r}). \label{kappa_r}
\end{eqnarray}
In general, real space non-condensate density and non-condensate correlations are nonlocal functions of two spatial points $\rho({\bf r}', {\bf r})$ and $\kappa({\bf r}', {\bf r})$. We only computed these quantities for ${\bf r} = {\bf r}'$ in this paper since these are the most physically accesible.  Measuring these quantities with ${\bf r} \neq {\bf r}'$ involves observing atomic correlations which is much more difficult than  photon correlations.

\subsection{Liouville space representation}

We next introduce the {\it Liouville space} notation\cite{spectroscopy,liouville} that will be used in the folllowing sections. One well-known example for which the Liouville space formalism is used is in the optical Bloch equations, where the $2 \times 2$
density matrix is recast as a 4 component vector, and the Liouvillian is
written accordingly as a $4 \times 4$ matrix superoperator. We rearrange the TIHFB equations Eqs. (\ref{z0dot}-\ref{kappa0dot}) by writing $\rho_{ij}$ and $\kappa_{ij}$ as vectors in Liouville space, and introduce the following set of $n^{2} \times n^{2}$ matrices (superoperators):
\begin{eqnarray}
{\cal H}^{(-)}_{ij, mn} & = & h^{(0)}_{im} \delta_{jn} - h^{(0)}_{nj}
\delta_{im}, \\
{\cal H}^{(+)}_{ij, mn} & = & h^{(0)}_{im} \delta_{jn} + h^{(0)}_{nj}
\delta_{im} + V_{ijmn}, \\
{\cal D}_{ij, mn} & = & \Delta^{(0)}_{im} \delta_{jn}, \\
{\cal D}_{ij, mn}^{\Delta} & = & -\Delta_{nj}^{*(0)} \delta_{im},
\end{eqnarray}
where the $n \times n$ matrices $h$ and $\Delta$ were defined in Eqs. (\ref{h}-\ref{Delta}). We further define the $n^{2} \times 1$ matrices
\begin{equation}
\Lambda^{\kappa}_{ij} =  \sum_{kl} V_{ijkl} z_{k}z_{l}.
\end{equation}
Using this notation, the TIHFB assume the form:
\begin{equation}
\left (
\begin{array}{cccccc}
{\cal H}_{z}^{(0)} & {\cal H}_{z*}^{(0)} & 0 & 0 & 0 & 0 \\
{\cal H}_{z*}^{(0)*} & {\cal H}_{z}^{(0)*} & 0 & 0 & 0 & 0 \\
0 & 0 & {\cal H}^{(-)} & {\cal D}^{\Delta} & 0 & {\cal D} \\
0 & 0 & -{\cal D}^{\Delta *} & {\cal H}^{(+)} & {\cal D} & 0 \\
0 & 0 & 0 & {\cal D}^{*} & {\cal H}^{(-)*} & {\cal D}^{\Delta *} \\
0 & 0 & {\cal D}^{*} & 0 & -{\cal D}^{\Delta} & {\cal H}^{(+)*}
\end{array}
\right ) \left (
\begin{array}{c}
\vec{z}^{(0)} \\
\vec{z}^{(0)*} \\
\vec{\rho}^{(0)} \\
\vec{\kappa}^{(0)} \\
\vec{\rho}^{(0)*} \\
\vec{\kappa}^{(0)*}
\end{array}
\right )
+ \left (
\begin{array}{c}
0 \\
0 \\
0 \\
\vec{\Lambda^{\kappa}} \\
0 \\
\vec{\Lambda^{\kappa}}^{*}
\end{array}
\right )
= 0. \label{SCTIHFB}
\end{equation}
Hereafter, we shall denote the zero'th order solution in Liouville space notation as $\vec{\psi}^{(0)}(t) \equiv [\vec{z}^{(0)}, \vec{z}^{(0)*},
\vec{\rho}^{(0)},\vec{\kappa}^{(0)},\vec{\rho}^{(0)*}, \vec{\kappa}^{(0)*}]^{T}$ and refer to the $2n(2n+1)$ by $2n(2n+1)$
matrix multiplying $\vec{\psi}^{(0)}$ as the zero'th order Liouville operator, ${\cal L}_{0}$.

Substituting the expansion
for $z_{i}$, $\rho_{ij}$, and $\kappa_{ij}$ [Eqs. (\ref{zexp} - \ref{kappaexp})] into Eqs. (\ref{zdot}-\ref{kappadot}),  we obtain to first order in $\eta$:
\begin{equation}
i\hbar \frac{d\vec{\psi}^{(1)}(t)}{dt} ={\cal L} \vec{\psi}^{(1)}(t) +
\zeta(t) .  \label{linear}
\end{equation}
Here $\vec{\psi}^{(1)} (t) =  [\vec{z}^{(1)}, \vec{z}^{(1)*},
\vec{\rho}^{(1)},\vec{\kappa}^{(1)},\vec{\rho}^{(1)*}, \vec{\kappa}^{(1)*}]^{T}$  i.e. a $2n(2n+1) \times 1$ vector with the variables in first order as its components. 
Also,
\begin{equation}
{\cal L} \equiv {\cal L}_{0} + {\cal L}_{1}
\end{equation}
where ${\cal L}$  is the total Liouvillian, ${\cal L}_{0}$ was introduced in Eq. (\ref{SCTIHFB}), and ${\cal L}_{1}$ and $\zeta(t)$, obtained by equating all first order terms in $\eta$ are given in Appendix \ref{L}. ${\cal L}$ is given by the sum of the original TIHFB
matrix ${\cal L}_{0}$ plus a perturbation ${\cal L}_{1}$. This perturbation induces a shift in the excitation frequencies, as will be shown below.

Adopting Liouville space notation, the position-dependent $n$th order solution  $\vec{\psi}^{(n)}({\bf r}, t)$ can be defined using the relations Eq. (\ref{z_r} - \ref{kappa_r}) and introducing a $2n(2n+1) \times 2n(2n+1)$ square matrix $\tilde{\Upsilon}({\bf r})$:
\begin{equation}
\vec{\psi} ^{(n)} ({\bf r}, t) \equiv \tilde{\Upsilon}({\bf r}) \vec{\psi}^{(n)} (t)
\end{equation}
where 
\begin{equation}
\tilde{\Upsilon}({\bf r}) = {\rm diag} \left [ \tilde{\phi}({\bf r}),
\tilde{\phi}^{*}({\bf r}), \Phi_{\rho}({\bf r}), \Phi_{\kappa}({\bf r}),
\Phi_{\rho}^{*}({\bf r}), \Phi_{\kappa}^{*}({\bf r}) \right ] . \label{upsilontilde}
\end{equation}
Here ``${\rm diag}[ \cdots ]$'' denotes that 
$\tilde{\Upsilon}({\bf r})$ is  a block diagonal square matrix made of $n \times n$ blocks $\tilde{\phi}({\bf r})$, $\tilde{\phi}^{*}({\bf r})$ and $n^2 \times n^2$ blocks $\Phi_{\rho}({\bf r}), \Phi_{\kappa}({\bf r}),
\Phi_{\rho}^{*}({\bf r}), \Phi_{\kappa}^{*}({\bf r})$.  $\tilde{\phi}({\bf r})$ is a diagonal matrix 
  with the $i$-th diagonal element given by the basis states $\phi_{i}({\bf r})$, and  $\Phi_{\rho}({\bf r})$ and $\Phi_{\kappa}({\bf r})$ are also diagonal matrices whose $ij$'th diagonal element  are given by 
$\left [ \Phi_{\rho}({\bf r}) \right ]_{ij,ij} = \phi^{*}_{i}({\bf r})\phi_{j}({\bf r})$, and $\left [ \Phi_{\kappa}({\bf r}) \right ]_{ij,ij} = \phi_{i}({\bf r})\phi_{j}({\bf r})$ respectively.  
The real space variables $z^{(n)} ({\bf r}, t)$,  $\rho^{(n)} ({\bf r}, t)$, and $\kappa^{(n)} ({\bf r}, t)$ are finally obtained by summing over the appropriate elements of the vector $\vec{\psi}^{(n)} ({\bf r}, t)$:
\begin{equation}
z^{(n)} ({\bf r}, t) = \sum_{i=1}^{n}  \vec{\psi}_{i}^{(n)} ({\bf r}, t) ,  \;\;\;\; \rho^{(n)} ({\bf r}, t) = \sum_{i = 2n+1}^{2n+n^2}  \vec{\psi}_{i}^{(n)} ({\bf r}, t) , \;\;\;\;  \kappa^{(n)} ({\bf r}, t) = \sum_{i = 2n + n^2+1}^{2n + 2n^2} \vec{\psi}_{i}^{(n)} ({\bf r}, t) .
\end{equation}

\section{The Response Functions}

The $n$'th order response function $K^{(n)}_{\alpha \rho}(t,t_1, \ldots t_n, {\bf r}, {\bf r}_1,\ldots {\bf r}_n)$ is defined by the relation:
\begin{equation}
\alpha^{(n)}({\bf r},t) = \int \int K^{(n)}_{\alpha \rho}(t,t_1, \ldots t_n,
{\bf r},{\bf r}_1,\ldots {\bf r}_n), V_{f}({\bf r}_1,t_1) \cdots
V_{f}({\bf r}_n,t_n) dt_1 \cdots dt_{n}  d{\bf r}_1 \cdots d^{3}{\bf r}_n, \label{K1r}
\end{equation}
where $\alpha^{(n)}({\bf r},t)$ with  $\alpha = z$, $\rho$, or $\kappa$ are position-dependent $n$'th order solutions. The $\rho$ subscript of $K^{(1)}_{\alpha \rho}$ indicate that this is  the response to an external field that couple to the atomic density. 

%It is noted that despite the assumption of local correlations of the non-condensate
%atoms, the interatomic interaction $V_{ijkl}$ is left general.

The $n$'th order susceptibility is defined as the Fourier transform of the response function to the frequency domain:
\begin{eqnarray}
\chi^{(n)}_{\alpha \rho}(\Omega, \Omega_{1}, \ldots, \Omega_{n}, {\bf r},{\bf r}_1,\ldots {\bf r}_n) & = &
\int_{0}^{\infty} dt \int_{0}^{\infty} dt_{1} \cdots 
dt_{n} K^{(n)}_{\alpha \rho}(t, t_{1}, \ldots, t_{n}, {\bf r},{\bf r}_1,\ldots {\bf r}_n)  \nonumber \\
& & \times \exp \left ( i \Omega t + i \Omega_{1} t_{1} + \cdots
+ i \Omega_{n} t_{n} \right ).
\end{eqnarray}

\subsection{The time-domain linear response function}

The solution of the matrix equation Eq. (\ref{linear}) is
\begin{equation}
\vec{\psi}^{(1)}(t) = \frac{1}{i\hbar }\int_{0}^{t}\exp \left[ -\frac{i}{%
\hbar }{\cal L}(t-t^{\prime })\right] \zeta (t^{\prime }) dt^{\prime } .
\end{equation}
The corresponding position-dependent solution can then be written:
\begin{eqnarray}
\vec{\psi}^{(1)}({\bf r}, t) &=&\frac{1}{i\hbar }\int d{\bf r}^{\prime } \int_{0}^{t} \tilde{\Upsilon}({\bf r}) {\cal U}(t-t^{\prime })\tilde{\Phi}(%
{\bf r}^{\prime })\vec{\psi}^{(0)}V_{f}({\bf r}^{\prime },t^{\prime
})dt^{\prime } \\
&=&\frac{1}{i\hbar }\int d{\bf r}^{\prime } \int_{0}^{t}\vec{K}^{(1)}_{\vec{\psi}}(t-t^{\prime }, {\bf r}, {\bf r}^{\prime
})V_{f}({\bf r}^{\prime },t^{\prime })dt^{\prime },
\label{K_1}
\end{eqnarray}
where $\tilde{\Upsilon}({\bf r})$ is as given in Eq. (\ref{upsilontilde}) and we have defined the $2n(2n+1) \times 2n(2n+1)$ matrices
\begin{eqnarray}
{\cal U}(t-t^{\prime }) & = & \theta(t-t') \exp \left[ -\frac{i}{\hbar }{\cal L}(t-t^{\prime }) \right ],   \label{U}  \\ 
\tilde{\Phi}({\bf r}) & = & {\rm diag}%
\left[ \Phi ({\bf r}),\Phi ^{\ast }({\bf r}),\Phi ^{(-)}(%
{\bf r}),\Phi ^{(+)}({\bf r}),\Phi ^{(-)\ast }({\bf r}),\Phi ^{(+)\ast }({\bf r})\right] , \label{phitilde}
\end{eqnarray}
with $\theta(t-t')$ being the Heaviside function and as discussed above, 
 the notation ``${\rm diag}[ \cdots ]$'' is used to  denote  
$\tilde{\Phi}({\bf r}^{\prime })$ as a $2n(2n+1) \times 2n(2n+1)$
block diagonal square matrix with the blocks consisting of  $n \times n$ square matrices having the $i$th row and $j$th column given by
\begin{equation}
\left[ \Phi ({\bf r})\right]_{ij}=\phi _{i}^{\ast }({\bf r})\phi _{j}({\bf r%
}),
\end{equation}
and $n^{2}\times n^{2}$ square matrices
\begin{equation}
\left[ \Phi ^{(\pm )}({\bf r})\right] _{ij,mn}=\phi _{i}^{\ast }({\bf r}%
)\phi _{m}({\bf r})\delta _{jn}\pm \phi _{n}^{\ast }({\bf r})\phi _{j}({\bf r%
})\delta _{im}.
\end{equation}

The function $\vec{K}^{(1)}_{\vec{\psi}} (t-t_1, {\bf r},{\bf r}_1)$ of Eq. (\ref{K_1}),
\begin{equation}
\vec{K}^{(1)}_{\vec{\psi}} (t-t_1, {\bf r},{\bf r}_1) \equiv \tilde{\Upsilon}({\bf r}) {\cal U}(t-t_1) \tilde{\Phi}({\bf r}_{1}) \vec{\psi}^{(0)}, \label{compactK1}
\end{equation}
may be viewed as  the linear response function for the position dependent vector $\vec{\psi}^{(1)}({\bf r}, t)$ i.e. for all the variables $z^{(1)}$, $\rho^{(1)}$ and $\kappa^{(1)}$ in the trap basis. The real space response functions for the condensate $z$, non-condensate density $\rho$, and the non-condensate correlation $\kappa$ are therefore given by summing over appropriate indices in $\vec{K}^{(1)}_{\vec{\psi}}(t-t_1, {\bf r},{\bf r}_1)$, using the relations Eq. (\ref{z_r}-\ref{kappa_r}):
\begin{eqnarray}
K^{(1)}_{z \rho}(t-t_1, {\bf r},{\bf r}_1) &=& \sum_{i=1}^{n} \vec{K}^{(1)}_{\vec{\psi}i}(t-t_1, {\bf r},{\bf r}_1), \label{K1z} \\
K^{(1)}_{\rho \rho}(t-t_1 , {\bf r}, {\bf r}_1) & = & \sum_{i = 2n + 1}^{2n +
n^2}  \vec{K}^{(1)}_{\vec{\psi}i}(t-t_1, {\bf r},{\bf r}_1),  \label{K1rho} \\
K^{(1)}_{\kappa \rho}(t-t_1, {\bf r},{\bf r}_1) & = & \sum_{i = 2n + n^2 + 1}^{2n
+ 2n^2}  \vec{K}^{(1)}_{\vec{\psi}i}(t-t_1, {\bf r},{\bf r}_1).  \label{K1kappa}
\end{eqnarray}

To analyze the physical significance of the response functions it will be useful to expand them in the basis of the eigenvectors $\vec{\xi}_{\nu}$ of matrix ${\cal L}$
\begin{equation}
{\cal L} \vec{\xi}_{\nu} = \omega_{\nu} \vec{\xi}_{\nu}, \;\;\;\; \nu = 1, 2, \ldots 2n(2n+1).
\end{equation}
We define the Green's function
\begin{equation}
G_{\nu}(t - t') = \theta(t-t') \exp \left[ -\frac{i}{\hbar }\omega_{\nu}(t-t^{\prime }) \right], \label{G_nu}
\end{equation}
and further introduce $\mu_{\nu}$, $\eta_{\nu}({\bf r}) $, and $\delta_{\nu} ({\bf r})$ as the expansion coefficients of the following vectors in the basis of eigenvectors $\vec{\xi}_{\nu}$
\begin{equation}
\vec{\psi}^{(0)} = \sum_{\nu = 1}^{2n(2n+1)} \mu_{\nu} \vec{\xi}_{\nu};  \;\;\;\;  \tilde{\Phi}({\bf r}) \vec{\xi}_{\nu} = \sum_{\nu = 1}^{2n(2n+1)}  \eta_{\nu}({\bf r}) \vec{\xi}_{\nu}; \;\;\;\; \tilde{\Upsilon} ({\bf r})  \vec{\xi}_{\nu} = \sum_{\nu = 1}^{2n(2n+1)}  \delta_{\nu} ({\bf r}) \vec{\xi}_{\nu}. \label{munudelta}
\end{equation}
We then have
\begin{equation}
\vec{K}^{(1)}_{\vec{\psi}}(t-t_1 , {\bf r}, {\bf r}_1) = \sum_{\nu} {\cal K}^{(1)}_{\nu \rho}(t,t' , {\bf r}, {\bf r}_1)   \vec{\xi}_{\nu}  \label{K1eigen}
\end{equation}
where
\begin{equation}
{\cal K}^{(1)}_{\nu \rho}(t,t' , {\bf r}, {\bf r}_1)  = \sum_{\nu', \nu''} \delta_{\nu} ({\bf r}) \eta_{\nu'} ({\bf r}_1) \mu_{\nu''} G_{\nu'}(t - t')  . \label{compactK1modes}
\end{equation}
Here, $G_{\nu}(t - t')$ is defined in Eq. (\ref{G_nu}),  $\mu_{\nu}$, $\eta_{\nu}({\bf r}) $, and $\delta_{\nu} ({\bf r})$ are defined as the expansion coefficients in Eq. (\ref{munudelta}), while the matrices $\tilde{\Phi}({\bf r})$ and $\tilde{\Upsilon} ({\bf r})$ are given by Eqs. (\ref{upsilontilde}) and (\ref{phitilde}).
Eqs. (\ref{K1eigen}-\ref{compactK1modes}) express the linear response function Eq. (\ref{compactK1}) as an expansion in quasiparticle modes.

\subsection{The frequency domain response function}

The linear susceptibility is defined as the Fourier Transform of the response function to the frequency domain:
\begin{equation}
\vec{\chi}^{(1)}( \Omega, \Omega_{1}, {\bf r},{\bf r}_1) = \int_{0}^{\infty} dt
\int_{0}^{\infty} dt_{1} \vec{K}^{(1)}( t-t_{1}, {\bf r},{\bf r}_1) \exp \left (
i \Omega t + i \Omega_{1} t_{1} \right ).
\end{equation}

It is possible to change variables $t - t_{1} \rightarrow \tau$ and define $\vec{\chi}^{(1)}(\Omega, {\bf r},{\bf r}_1)$ since $\vec{K}^{(1)}( t-t_{1}, {\bf r},{\bf r}_1)$ only depends on the time difference, $t-t_{1}$. Below, we shall keep the notation $\vec{K}^{(1)}( \Omega, \Omega_{1}, {\bf r},{\bf r}_1)$ i.e. a function of both frequency of the signal, $\Omega$, and frequency of perturbation, $\Omega_1$, rather than $\vec{K}^{(1)}( \Omega, {\bf r},{\bf r}_1)$, in line with Bloembergen's notation\cite{spectroscopy}.

In order to evaluate the Fourier Transform, we note that it follows from causality
\begin{eqnarray}
{\cal U}(t) & = & - \frac{1}{2 \pi i} \int_{-\infty}^{\infty} d \omega
\frac{1}{\omega - {\cal L} + i \epsilon} \exp(- i \omega t) \\
& = & \int_{-\infty}^{\infty} d \omega
{\cal U}(\omega) \exp(- i \omega t).  \label{FT}
\end{eqnarray}

Substituting Eq. (\ref{FT}) into the expression for $\vec{K}^{(1)}_{\vec{\psi}}( t-t_{1}, {\bf %
r},{\bf r}_1)$ Eq. (\ref{compactK1}), and taking the Fourier Transform, we obtain a vector:
\begin{eqnarray}
\vec{\chi}^{(1)}_{\vec{\psi}}( \Omega, \Omega_{1}, {\bf r},{\bf r}_1) & = & \frac{1}{2 \pi i}
\int_{-\infty}^{\infty} d \omega \int_{0}^{\infty} dt \/
dt_{1} \tilde{\Upsilon}({\bf r}) {\cal U}(\omega) \exp(- i \omega t + i \omega t_{1}) \exp
\left ( i \Omega t + i \Omega_{1} t_{1} \right ) \tilde{\Phi}({\bf r}_{1})
\vec{\psi}^{(0)} \\
& = & \frac{1}{2 \pi i} \int_{-\infty}^{\infty} d \omega \tilde{\Upsilon}(%
{\bf r}) {\cal U}(\omega) \tilde{\Phi}({\bf r}_{1}) \vec{\psi}%
^{(0)}\delta(\Omega - \omega) \delta(\Omega_{1} + \omega) .
\end{eqnarray}

This implies that  $\Omega = -\Omega_{1}$, and we have the susceptibility in vector form
\begin{equation}
\vec{\chi}^{(1)}_{\vec{\psi}}(-\Omega_{1}; \Omega_{1}, {\bf r},{\bf r}_1) = \tilde{\Upsilon}({\bf r%
}) {\cal U}(\Omega_1) \tilde{\Phi}({\bf r}_{1}) \vec{\psi}^{(0)}. \label{freqdom_K1}
\end{equation}
The $z$, $\rho$ and $\kappa$ susceptibilities are obtained by summing over the appropriate elements in the vector $\vec{\chi}^{(1)}_{\vec{\psi}}(-\Omega; \Omega, {\bf r},{\bf r}_1)$:
\begin{eqnarray}
\chi^{(1)}_{z \rho}(-\Omega; \Omega, {\bf r},{\bf r}_1) &=& \sum_{i =1}^{n}  \vec{\chi}^{(1)}_{\vec{\psi}i}(-\Omega; \Omega, {\bf r},{\bf r}_1)   \label{freqdom_K1z} \\
\chi^{(1)}_{\rho \rho}(-\Omega; \Omega, {\bf r},{\bf r}_1) &=& \sum_{i = 2n+1}^{2n+n^2}  \vec{\chi}^{(1)}_{\vec{\psi}i}(-\Omega; \Omega, {\bf r},{\bf r}_1)   \label{freqdom_K1rho}
\\
\chi^{(1)}_{\kappa \rho}(-\Omega; \Omega, {\bf r},{\bf r}_1) &=&
\sum_{i =  2n + n^2 + 1}^{2n+2n^2}   \vec{\chi}^{(1)}_{\vec{\psi}i}(-\Omega; \Omega, {\bf r},{\bf r}_1)  .  \label{freqdom_K1kappa}
\end{eqnarray}

Transforming the trap basis to the basis of eigenstates $\vec{\xi}_{\nu}$ as before, one has:
\begin{equation}
\vec{\chi}^{(1)}_{\vec{\psi}}(-\Omega; \Omega, {\bf r},{\bf r}_1) = \sum_{\nu} {\cal K}_{\nu \rho}^{(1)}(-\Omega; \Omega_, {\bf r},{\bf r}_1)  \vec{\xi}_{\nu}, \label{freqdom_K1sum}
\end{equation}
where
\begin{equation}
{\cal K}_{\nu \rho}^{(1)}(-\Omega; \Omega, {\bf r},{\bf r}_1) = \sum_{\nu' \nu''} \frac{\delta_{\nu}({\bf r}) \eta_{\nu'}({\bf r}_1) \mu_{\nu''}}{\Omega -\omega_{\nu'}  + i
\epsilon} ,  \label{freqdom_K1modes}
\end{equation}
and $\delta_{\nu}({\bf r})$,  $\eta_{\nu}({\bf r}_1)$,  and $\mu_{\nu}$ are as defined in Eq.(\ref{munudelta}).

As was done in Section III A for the time domain, we have recast the linear susceptibility in two forms:  matrix form of Eqs. (\ref{freqdom_K1}) and an expansion in modes of Eq. (\ref{freqdom_K1sum}-\ref{freqdom_K1modes}).

\section{Numerical simulations for contact potential}

\subsection{Zero'th order solution (TIHFB) and the frequency shifts}

So far, all our results hold for a general pairwise interatomic interaction potential. In the following numerical calculations, we approximate the interatomic potential $V({\bf r}-{\bf r^{\prime}})$ in Eq. (\ref{Vijkl}) by a contact potential, as is standard in BEC applications:
\begin{equation}
\begin{array}{c@{\hspace{1cm}}c}
{\displaystyle V({\bf r}-{\bf r^{\prime}}) \rightarrow U_{0} \delta({\bf r}-%
{\bf r^{\prime}}), } & {\displaystyle U_{0} = \frac{4\pi\hbar^{2}a}{m}, }
\end{array}
\label{contactpotential}
\end{equation}
where $a$ is the {\em s}-wave scattering length and $m$ is the atomic mass. This approximation may be justified since the wave functions at ultracold temperatures have very long wavelengths compared to the range of interatomic potential. This implies that details of the interatomic potential become unimportant and the potential may be approximated by a  contact potential.
The tetradic matrices $V_{ijkl}$ are then simply given by:
\begin{equation}
V_{ijkl} = \frac{4\pi\hbar^{2}a}{m} \int \phi^{*}_{i}({\bf r})\phi^{*}_{j}(%
{\bf r})\phi_{k}({\bf r})\phi_{l}({\bf r}) d{\bf r}.
\end{equation}

We assume a 2000 atom one dimensional condensate in a harmonic trap. The parameters used for our numerical calculations are:  $U_0 = \frac{4\pi\hbar^{2}a}{m} = 0.01$, and  temperatures $0 \hbar \omega_{\rm trap}/k_B$ and $10 \hbar \omega_{\rm trap}/k_B$ where $\omega_{\rm trap}$ is the trap frequency and $k_B$ is the Boltzmann constant. We used 256 grid points for position and the basis set of $n=5$ states. We keep the trap units throughout.

We have followed the prescription of Griffin for solving the TIHFB for $\vec{\psi}^{(0)}$ in terms of the Bogoliubov-de Gennes Equations which are obtained from Eq. (\ref{HR}) by transforming to real space and using the contact interatomic potential\cite{Griffin}. In the Appendix we show the equivalence between the Bogoliubov-de Gennes Equations and the matrix $H^{(0)}$, and  summarize the numerical procedure.
The eigenvalues of the non-Hermitian matrix ${\cal L}$ required for computing the response functions were calculated using the Arnoldi algorithm\cite{Arnoldi}.

The eigenvalues of the Liouvillian are trasition frequencies rather than state energies.  The frequencies come in pairs of positive and negative frequencies; this indicates that the Liouvillian may be mapped onto a harmonic oscillator space for which there are always positive and negative frequency solutions. The eigenvalues of ${\cal L} \equiv {\cal L}_0 + {\cal L}_1$ are shifted with respect to the corresponding eigenvalues of ${\cal L}_{0}$ (the TIHFB equations). We list some of the representative eigenvalues, namely the lowest few positive eigenvalues of ${\cal L}_{0}$ and ${\cal L}$ in Table \ref{eigenlist} for both zero and nonzero temperatures. Similar frequency shifts were noted by Giorgini\cite{Giorgini}. Physically it is easy to understand how the TDHFB frequencies  may be shifted from the TIHFB: a dynamical system contains the effect of the interacting condensate and non-condensate  atoms  which, by definition, is not present in the equilibrium system. This is analogous to optical excitations of fermions where the time-dependent Hartree-Fock theory shows excitonic shifts which are lacking by the time-independent Hartree-Fock states\cite{TDHF}.

\subsection{Time domain response}

The linear response functions were calculated by substituting the numerical solution $\vec{\psi}^{(0)}$ evaluated at zero and finite temperatures into Eq. (\ref{compactK1}) together with Eqs. (\ref{K1z}-\ref{K1kappa}).  Eq. (\ref{compactK1}) is a matrix multiplication of $2n(2n+1) \times 1$ vector $\vec{\psi}^{(0)}$ with $2n(2n+1) \times 2n(2n+1)$ matrices $\tilde{\Upsilon}$, ${\cal U}(t)$, and $\tilde{\Phi}$ which are defined in Eqs. (\ref{phitilde}) and (\ref{upsilontilde}).
$\tilde{\Phi}$ and $\tilde{\Upsilon}$ are constructed in terms of the harmonic oscillator basis states which are calculated numerically from the recursive formula that involves the Gaussian function multiplying the Hermite polynomials\cite{Arfken}. The matrix ${\cal U}(t)$ was calculated using a MATLAB function that uses the Pad\'{e} approximation for matrix exponentiation\cite{Golub}.

We first present the dependence of the linear response function on ${\bf r}$ and ${\bf r}^{\prime}$ at fixed times $t-t'$. This gives a snapshot of the position-dependent correlations across the condensate. Such dependence is important since the experimentally produced condensates are mesoscopic in size; in contrast, the dipole approximation usually applies in optical spectroscopy and consequently the spatial dependence of the response is not observable.

In Figs \ref{K1rrtT0} we display the real space response functions at times $t - t' = 0 /\omega_{\rm trap}, 7.2/\omega_{\rm trap}, 15.7/\omega_{\rm trap}$ at zero temperature. Fig. \ref{K1rrt} shows the corresponding finite temperature results. Physically, $t'$ and ${\bf r}'$ are respectively the time and postion at which the external perturbation is applied and $t$ and  ${\bf r}$ are the corresponding coordinates at which the measurement is made. We found that for longer times $t-t' > 5 \pi/\omega_{\rm trap}$, the plot maintains generally the same shape as the third column of figures; it may therefore be possible to experimentally observe the correlations at longer times.
At zero temperature, the correlation attains this stable shape faster than at finte temperature.

To model a uniform perturbation applied across the condensate, we have integrated the zero temperature response function over ${\bf r}'$ and plotted its absolute value i.e. $|\int K^{(1)}({\bf r}, {\bf r}' , t-t_1) d {\bf r}'|$ vs. time $t-t'$ in Fig. \ref{K1rtT0}. The response functions $K_{z \rho}$, $K_{\rho \rho}$, $K_{\kappa \rho}$ grow exponentially within a relatively short time span of around $5 \pi/\omega_{trap}$, so that the details of the structure for earlier times up to $\sim 2 \pi/\omega_{trap}$  are not visible in the plot.
This rapid growth is shown more clearly in the right hand column which depicts the response functions integrated over both ${\bf r}$ and ${\bf r}'$, i.e. $K^{(1)}(t-t_1)$. The figure shows the real and imaginary parts as well as the absolute values of $K^{(1)}(t-t_1)$. The response function grows rapidly by around an order of magnitude over the time scale $\sim 3 \pi/\omega_{trap}$. 
 The response function keeps growing over time since TDHFB does not have any dissipative term; this should be a reasonable model for BEC in the collisionless regime.  It also shows that, even in the absence of a dissipative term, it takes some time after the initial impulse at $t'=0$  before the effect of the force is reflected appreciably in the response functions. From the plots we note that a mechanical force applied on the condensate can be seen to  ``generate'' noncondensate atoms and anomalous correlations even at zero temperature.

  The calculations of Fig. \ref{K1rtT0} are repeated at finite temperature in Fig. \ref{K1rt}. The  main difference is the smaller magnitude of $K_{z \rho}$, $K_{\rho \rho}$ and $K_{\kappa \rho}$. The fact that the BEC is less responsive at finite temperature may be attributed to the fact that the condensate to non-condensate interaction is greater at finite temperatures where additional collisions shield the effect of the applied perturbation.

\subsection{Frequency domain response}

The susceptibilities were computed using Eq. (\ref{freqdom_K1}) in conjunction with Eqs. (\ref{freqdom_K1z}-\ref{freqdom_K1kappa})  Eq. (\ref{freqdom_K1}) is a matrix multiplication of $2n(2n+1) \times 1$ vector $\vec{\psi}^{(0)}$ with $2n(2n+1) \times 2n(2n+1)$ matrices $\tilde{\Upsilon}$, ${\cal U}(\omega)$, and $\tilde{\Phi}$.
The matrix ${\cal U}(\omega)$ is calculated as follows:
\begin{equation}
{\cal U}(\omega) = \frac{1}{\omega - {\cal L} + i \epsilon} = \sum_{\nu} \frac{\xi_{\nu}\zeta_{\nu}^{\dagger}}{\omega - \omega_{\nu} + i \epsilon}
\end{equation}
where $\xi_{\nu}$ is the right eigenvalue of ${\cal L}$  with eigenvalues $\omega_{\nu}$ (${\cal L} \xi_{\nu}  =  \omega_{\nu} \xi_{\nu}$) and $\zeta_{\nu}$ are the left eigenvectors such that $\sum_{\nu}  \xi_{\nu}\zeta_{\nu}^{\dagger} =  \openone$. The eigenvalues $\omega_{\nu}$ of ${\cal L}$ were calculated using the Arnoldi algorithm\cite{Arnoldi}.

To clearly display the resonance structure,  we present in Fig. \ref{K1spec} the absolute value of zero and finite temperature linear response  integrated over ${\bf r}$ and ${\bf r}_1$ in the frequency domain, $|\int  \chi^{(1)}_{\alpha \rho}(-\Omega, \Omega, {\bf r}, {\bf r}' ) d {\bf r} d {\bf r}'|$, $\alpha = z, \rho, \kappa$ on a  logarithmic scale.  The eigenvalues of ${\cal L}$ and hence the resonant frequencies come in positive and negative pairs; we present only the positive frequencies since the function is symmetric about the zero frequency.  We note that low frequency resonances are dominant; this may explain the absence of any oscillatory features in the time domain plots.

Similar to the time domain calculations of Figs. \ref{K1rrtT0} and \ref{K1rrt}, we show in Fig. \ref{K1rrwT0} the linear response,  $\chi^{(1)}(-\Omega,\Omega, {\bf r}, {\bf r}_1)$, at zero temperature as a function of ${\bf r}$ and  ${\bf r}_1$  for 3 different frequencies.  These frequencies represent the resonant frequency corresponding to the strongest peak for the condensate at $\Omega = 0 \omega_{\rm trap}$, an off-resonant frequency at $\Omega = 0.25\omega_{\rm trap}$, and the resonant frequency corresponding to the second highest peak for the condensate at $\Omega = 0.46\omega_{\rm trap}$. In Fig. \ref{K1rrw} we repeat the calculations for a finite temperature, and the frequencies represent the resonant frequency corresponding to the strongest peak for the condensate at $\Omega = 0.63\omega_{\rm trap}$, an off-resonant frequency at $\Omega = 1.25\omega_{\rm trap}$, and the resonant frequency corresponding to the second highest peak for the condensate at $\Omega = 1.7\omega_{\rm trap}$. Similar to the time domain plots, the frequency domain response is strongly position dependent. Off-resonant frequencies give a more complicated pattern.

The absolute values of the zero and finite temperature  $\chi^{(1)}(-\Omega; \Omega, {\bf r}, {\bf r}')$ integrated  over ${\bf r}'$, $| \int \chi^{(1)}(-\Omega; \Omega, {\bf r}, {\bf r}') d {\bf r}'|$,  are plotted  as a function of $\Omega$ in Figs. \ref{K1rwT0} and \ref{K1rw}. This represents the response to a spatially uniform external perturbation.
Since the resonance peaks vary vastly in strength we also present a contour plot in which all the intensities have been normalized to 1.

\section{Discussion}

We have applied the systematic formalism of nonlinear spectroscopy  to calculate the response functions and susceptibilities of  BEC to a  mechanical force coupled to atomic density for the condensate, noncondensate density, and noncondensate correlations. Since our results hold for an arbitrary external perturbation, it will be interesting to investigate the effects of specifically taylored external force e.g. impulsive or ``continuous wave'' perturbations.  The contour plots of the response functions and susceptibilities  may be used for the design of experiments involving applications of mechanical forces to a condensate. For instance, by carefully specifying the shape of the external potential $V_{f}({\bf r}, t)$, one may effectively cancel out the response of, say, the noncondensate atoms. This would provide new insights into the dynamics of the condensate and noncondensate interactions.  Our results show the time delay between the application of a mechanical force at $t'=0$ and the build up of response. We further note that the condensate and noncondensate atoms as well as the noncondesate correlations respond to the external mechanical force.

The response functions computed in this paper offer a practical way to solve the TDHFB numerically with modest computational effort. Still, one potential bottleneck is the computational cost required to diagonalize a large matrix. Krylov space techniques such as the Arnoldi algorithm used in the paper considerably reduces that cost\cite{Arnoldi}.  As our matrices were only $110 \times 110$ in these simulations of 1 dimensional condensate the entire set of eigenvalues could be calculated directly. For larger matrices, the algorithm only gives the lowest few eigenvalues. This should be sufficient to describe realistic experiments such as 3 dimensional asymmetric trap holding up to a million atoms.

The damping of excitations is not included in TDHFB.  In Ref. \cite{Giorgini} Landau and Beliaev damping were introduced by calculating the imaginary part of the self energy. 
%Landau damping is the process in which one quantum of oscillation $\hbar \omega$ interacts with %a quasiparticle excitation $E_{i}$ to give another quasiparticle excitation with energy $E_{j} %= E_{i} + \hbar \omega$, while in Beliaev damping, a quantum of oscillation of energy $\hbar 
%\omega$ is absorbed to give two quasiparticle excitations such that the total energy matches 
%$\hbar \omega$: $\hbar \omega = E_{i} + E_{j}$. 
The theory of damping of oscillations in BEC is currently far from conclusive and requires further investigation. This has motivated various authors to try and extend the  HFB theory\cite{Prou1,Fedichev_Shlyapnikov,Pitaevskii_Landaudamping,ZNG,Zaremba}.  It should be noted that there is currently no all-encompassing theory of BEC that explains all  observed phenomena. In addition, as noted by Leggett\cite{leggett}, the validity of various approximations made in some of the existing theories of strongly nonequilibrium dynamics of BEC (e.g. kinetics of the condensation process, the damping of collective excitations, and the decay of vortex states) is not entirely clear. Equations such as ZNG constitute a very important contribution. The TDHFB equations do constitute a systematic and consistent description of trapped atomic BEC at finite temperatures in the collisionless regime,  just as the Gross-Pitaevskii Equation is  valid near zero temperature  and  in the future it should be possible to observe directly the effects of anomalous correlations.

Most current work on BEC excitations deals only with the linear response; however nonlinear effects should be observable with stronger perturbations. Just as in standard nonlinear optics, the $n$'th order response functions are expected to be most valuable for characterizing BEC in the context of  matter-wave nonlinear optics. Our formalism allows the calculation of nonlinear susceptibilities which may be used to probe finite temperature condensates by four wave mixing\cite{Band}. Nonlinear response functions will be computed in a forthcoming work.  Other possible future applications include detailed study of atom optics at finite temperatures, and superchemistry that involves the study of the formation of molecules from mesoscopic BEC matter-waves\cite{future}.

\acknowledgments
The support of NSF Grant No. CHE-0132571 is gratefully acknowledged.

\appendix

\section{Equilibrium TIHFB Solution}

%Substituting Eqs. (\ref{zexp}-\ref{kappaexp}) into Eqs. (\ref{zdot}- \ref{kappadot}) and equating %zero'th power in $\eta$, it is found that  $z^{(0)}$ obeys 

The TIHFB is given by the coupled equations:
\begin{equation}
{\cal H}_{z}^{(0)} z^{(0)} + {\cal H}_{z*}^{(0)} z^{*(0)} = 0,  \label{NLSE}
\end{equation}
\begin{equation}
\tilde{\gamma} [H^{(0)}\tilde{\gamma}, \tilde{\gamma} R^{(0)}] = 0, \label{BRTIHFB}
\end{equation}
where Eq. (\ref{NLSE}) is the time-independent GPE with ${\cal H}_{z}^{(0)}$ and  ${\cal H}_{z*}^{(0)}$ defined respectively in Eqs. (\ref{h_z}) and (\ref{h_z_star}), and
the matrices $R^{(0)}$, $H^{(0)}$,  $\tilde{\gamma}$ of Eq. (\ref{BRTIHFB}) are as defined in Eq. (\ref{HR}).

The equilibrium solution to TIHFB is obtained variationally. 
%For convenience, Eqs. (\ref{NLSE}) and (\ref{BRTIHFB}) are reproduced below:
%\begin{eqnarray}
%{\cal H}_{z}^{(0)} z^{(0)} + {\cal H}_{z*}^{(0)} z^{*(0)} & = & 0, \label{AppNLSE} %\\
%\tilde{\gamma} [H^{(0)}\tilde{\gamma}, \tilde{\gamma} R^{(0)}] & = & 0. 
%\label{AppBRTIHFB}
%\end{eqnarray} 
Eq. (\ref{NLSE}) is solved for $z^{(0)}$ by using a numerical optimization routine to minimize the energy functional $E = \langle  \hat{H}   \rangle$ with respect to $z^{(0)*}$. The energy functional is found by applying the thermal Wick's theorem  to the Hamiltonian $\hat{H}$. For the Wick theorem, Eq. (\ref{Wick3}), to hold the state of the system is assumed to be described by a trial statistical density matrix of the form
\begin{equation}
D = \frac{e^{-\beta K}}{Z_0}, \;\;\;\;\;\; Z_0  = {\rm Tr} e^{-\beta K}
\end{equation}
where the operator $K$ is taken to be quadratic in the annihilation/creation operator for the non-condensate atoms, $c_i/c^{\dagger}_i$ defined as $c_i  \equiv a_i - z_i$/$c^{\dagger}_i  \equiv a^{\dagger}_i - z^{*}_i$:
\begin{equation}
K = \frac{1}{2} \sum_{ij} [ h^{(0)}_{ij} (c^{\dagger}_{i} c_{j} + c_{j} c^{\dagger}_{i} ) + \Delta^{(0)}_{ij}  c^{\dagger}_{i}c^{\dagger}_{j} + \Delta_{ij}^{(0)*}  c_{i}c_{j}],
\end{equation}
with
\begin{equation}
\rho^{(0)}_{ij} = {\rm Tr} D c^{\dagger}_{j}c_{i}, \;\;\;\;\; \kappa^{(0)}_{ij} = {\rm Tr} D c_{i}c_{j}. 
\end{equation}
Then one obtains for the energy $E = \langle H \rangle$:
\begin{eqnarray}
E & = & \sum_{ij} (H_{ij} - \mu) \left [ z^{(0)*}_{i} z^{(0)}_{j} +  \rho^{(0)}_{ij} \right ]  + \frac{1}{2}\sum_{ijkm} V_{ijkm} \left [ z^{(0)*}_{i} z^{(0)*}_{j} z^{(0)}_{k} z^{(0)}_{m} + 4z_{i}^{(0)* } z^{(0)}_{k}  \rho^{(0)}_{jm}  \right. \nonumber \\
 & + &  \left. z_{i}^{(0)*} z_{j}^{(0)*} \kappa^{(0)}_{km}  + \kappa_{ij}^{(0)*} z^{(0)}_{k} z^{(0)}_{m}  +  2 \rho^{(0)}_{ik} \rho^{(0)}_{jm}  + \kappa_{ij}^{(0)*} \kappa^{(0)}_{km} \right ] .
\label{Energy} %+ \sum_{i} \lambda_{i} z^{*}_{i} + \lambda^{*}_{i} z_{i}
\end{eqnarray}
It is easy to show that the local minimum for $E$ obtained by setting its derivative with respect to $z^{(0)*}$ to zero satisfies Eq. (\ref{NLSE}).

In addition, $R^{(0)}$ is found by minimizing the thermodynamic potential for a system of bosons in thermal equilibrium.  In Ref. \cite{Blaizot_Ripka}, the generalized density matrix $R^{(0)}$ in thermal equilibrium, assuming a grand canonical form for the density matrix is shown to be given by:
\begin{equation}
R^{(0)} =  \frac{1}{\exp(\tilde{\gamma} H^{(0)}/kT) - 1} \tilde{\gamma} . \label{eqmR}
\end{equation}
It may be shown straightforwardly that the generalized density matrix of Eq. (\ref{eqmR}) obeys Eq. (\ref{BRTIHFB}), and is therefore a stationary solution. The proof requires using the property $\tilde{\gamma}^2 = \openone$, and the fact that a matrix $A$ commutes with a function of $A$, $f(A)$.
The minimization of the thermodynamic potential also gives the relation\cite{Blaizot_Ripka} 
\begin{equation}
\frac{1}{2} H^{(0)}_{ij} = \frac{\partial E}{\partial R^{(0)}_{ij}}. \label{minH0}
\end{equation}
Using Eq. (\ref{Energy})  and the definition of $R^{(0)}$ given in Eq. (\ref{HR}), Eq. (\ref{minH0}) implies  
\begin{eqnarray}
h^{(0)}_{ij} & = & \frac{\partial E}{\partial \rho^{(0)}_{ij}} = H_{ij} -\mu  + 2 \sum_{kl} \langle ik| \hat{V} | lj \rangle \left [
z_{k}^{(0)*}z^{(0)}_{l} + \rho^{(0)}_{lk} \right ], \\
\Delta^{(0)}_{ij} & = & \frac{\partial E}{\partial \kappa^{(0)*}_{ij}} = \sum_{kl} \langle ij| \hat{V} | kl \rangle \left[ z^{(0)}_{k} z^{(0)}_{l}
+ \kappa^{(0)}_{kl} \right].
\end{eqnarray}
These variational results satisfy Eqs. (\ref{h}-\ref{Delta}), derived using the Heisenberg equations of motion.  
%\begin{eqnarray}
%\tilde{\gamma} [H^{(0)} \tilde{\gamma}, \tilde{\gamma} R^{(0)}]   & = &  %\tilde{\gamma} H^{(0)} \frac{1}{\exp(\tilde{\gamma} H^{(0)}/kT) - 1 } %\tilde{\gamma}
%- \frac{1}{\exp(\tilde{\gamma} H^{(0)}/kT) - 1 } \tilde{\gamma} H^{(0)} %\tilde{\gamma} \\
%& = & [\tilde{\gamma} H^{(0)},  \frac{1}{\exp(\tilde{\gamma} H^{(0)}/kT) - 1 }] %\tilde{\gamma} =  0 ,
%\end{eqnarray}
%\begin{equation}
%\tilde{\gamma} [H^{(0)} \tilde{\gamma}, \tilde{\gamma} R^{(0)}]  =  0 ,
%\end{equation}

We now discuss the self-consistent solution to TIHFB. The solution to TIHFB involves simultaneous minimization of Eq. (\ref{Energy}) coupled to Eq. (\ref{eqmR})  where $H^{(0)}$ is as defined in Eqs.  (\ref{h}-\ref{Delta}) and (\ref{HR}).
Since $H^{(0)}$ is itself a function of $R^{(0)}$, the solution is found iteratively. 
In order to evaluate $R^{(0)}$, Eq. (\ref{eqmR}), we need to diagonalize the matrix $\tilde{\gamma} H^{(0)}$. The specific form of the matrix $\tilde{\gamma} H^{(0)}$ implies that its eigenvalues and eigenvectors come in pairs. Letting $\tilde{V}^{n}$ and $\tilde{W}^{n}$ denote right eigenvectors belonging to the eigenvalues $\pm E_{n}$,
\begin{equation}
\tilde{\gamma} H^{(0)}\tilde{V}^{n} = E_{n}\tilde{V}^{n}, \;\;\;\;\; \tilde{\gamma} H^{(0)}\tilde{W}^{n} = -E_{n}\tilde{W}^{n}, \;\;\;\;\; E_{n} >  0, \label{eigenvalue}
\end{equation}
where the normalisation and closure relations of the eigenvectors are:
\begin{equation}
\tilde{V}^{n\dagger}\tilde{\gamma}\tilde{V}^{m} = \delta_{mn}, \;\;\;\;\; \tilde{W}^{n\dagger}\tilde{\gamma}\tilde{W}^{m} = -\delta_{mn}, \;\;\;\;\; \tilde{V}^{n\dagger}\tilde{\gamma}\tilde{W}^{m} = 0 \label{VWnorm}
\end{equation}
and
\begin{equation}
\sum_{n \geq 0} \left ( \tilde{V}^{n}V^{n \dagger}\tilde{\gamma} - \tilde{W}^{n}\tilde{W}^{n \dagger}\tilde{\gamma}  \right ) = 1.
\end{equation}
It is noted that he eigenvectors $\tilde{V}^{n}$ and $\tilde{W}^{n}$ have the following structure:
\begin{equation}
\tilde{V}^{n} = \left (
\begin{array}{c}
U^{n} \\
V^{n}
\end{array} \right ), \;\;\;\;\;\;
\tilde{W}^{n} = \left (
\begin{array}{c}
V^{n*} \\
U^{n*}
\end{array} \right ),
\end{equation}
and given a right eigenvector $\tilde{V}^{n}$,  the left eigenvector belonging to the eigenvalue $E_n^{*}$ is: $\bar{V}^{n} = ( U^{n*}, - V^{n*})$, and the left eigenvector belonging to the eigen value $-E_n^{*}$ is: $\bar{W}^{n} = ( V^{n}, -U^{n})$.

Using the closure relationship, $R^{(0)}$ of Eq. (\ref{eqmR})
may be  written  as:
\begin{equation}
R^{(0)}_{ij} =  \sum_{m>0} \left [ \tilde{V}_{i}^{m}\tilde{V}_{j}^{m*}  +
\tilde{W}_{i}^{m}\tilde{W}_{j}^{m*}  \right ] \bar{n}_{m}  + \sum_{m>0} \tilde{W}_{i}^{m}\tilde{W}_{j}^{m*}  \label{VWR}
\end{equation}
with $\bar{n}_m = \left [ \exp(E_{m}/kT) - 1  \right ]^{-1}$. The possible zero energy states are ignored, as indicated in the summation over $m > 0$.
From this expression, one can calculate the matrices $\rho$ and $\kappa$  from the explicit form of $R$ given in Eq. (\ref{HR}).

The iteration process consists of the follwoing steps\cite{Blaizot_Ripka}:
\begin{enumerate}
\item Make an initial guess at ${\cal H}_{z}^{(0)}$ and set $\rho =\kappa = 0$
\item Solve Eq. (\ref{NLSE}) which determines $\mu$ and $z$. Normalize $z$ according to:
\begin{equation}
N = \sum_{i} |z_{i}|^2 + {\rm tr} \rho
\end{equation}
where $N$ is the total number of atoms and ${\rm tr}$ denotes the trace.
\item Solve the eigenvalue problem Eq. (\ref{eigenvalue}) and normalize the eigenvctors according to Eq. (\ref{VWnorm})
\item Calculate $R$ from Eq. (\ref{VWR}) and deduce the matrices $\rho$ and $\kappa$.
\item Calculate the fields ${\cal H}_{z}^{(0)}$ and  ${\cal H}_{z*}^{(0)}$ and return to step 2.
\item Stop the iteration when two successive iterations yield the same values of $z$, $\rho$ and $\kappa$ to the desired accuracy.
\end{enumerate}

\section{The ${\cal L}_{1}$ matrix} \label{L}

In Eq. (\ref{linear}), ${\cal L} \equiv {\cal L}_{0} + {\cal L}_{1}$. The $2n(2n+1) \times 2n(2n+1)$ matrix ${\cal L}_{1}$ is defined as follows:
\begin{equation}
{\cal L}_{1} = \left (
\begin{array}{cccccc}
{\cal V}^{zz1} & {\cal V}^{zz2} & {\cal V}^{z1} & {\cal V}^{z2} & 0 & 0 \\
{\cal V}^{zz2*} & {\cal V}^{zz1*} & 0 & 0 & {\cal V}^{z1*} & {\cal V}^{z2*}
\\
{\cal V}^{\rho z1} & {\cal V}^{\rho z2} & {\cal W}^{\rho h} & {\cal W}%
^{\kappa \Delta} & 0 & {\cal W}^{\kappa \Delta \dagger} \\
{\cal V}^{\kappa z1} & {\cal V}^{\kappa z2} & {\cal W}^{\kappa h} & {\cal W}%
^{\rho \Delta} & {\cal W}^{\kappa h \dagger} & 0 \\
{\cal V}^{\rho z2*} & {\cal V}^{\rho z1*} & 0 & ({\cal W}^{\kappa \Delta
\dagger})^{*} & ({\cal W}^{\rho h})^{*} & ({\cal W}^{\kappa \Delta})^{*} \\
{\cal V}^{\kappa z2*} & {\cal V}^{\kappa z1*} & ({\cal W}^{\kappa h
\dagger})^{*} & 0 & ({\cal W}^{\kappa h})^{*} & ({\cal W}^{\rho \Delta})^{*}
\end{array}
\right ),
\end{equation}
where the set of $n \times n$ submatrices ${\cal V}^{zz1}$ and ${\cal V}^{zz2}$, $n \times n^{2}$ submatrices  ${\cal V}^{z1}$ and ${\cal V}^{z2}$,
$n^2 \times n$ submatrices
${\cal V}^{\rho z1}$,  ${\cal V}^{\rho z2}$,  ${\cal V}^{\kappa z1}$, and  ${\cal V}^{\kappa z2}$, and $n^{2} \times n^{2}$ component submatrices
 ${\cal W}^{\rho h}$, ${\cal W}^{\rho \Delta}$, ${\cal W}^{\kappa h}$, and ${\cal W}^{\kappa \Delta}$ of ${\cal L}_{1}$ are given as follows:
\begin{eqnarray}
{\cal V}^{zz1}_{i,l} & = & \sum_{kr} V_{iklr} z^{*(0)}_{k}z^{(0)}_{r} \;\;\;\;\; {\cal V}^{zz2}_{i,k}  =  \sum_{lr} V_{iklr} z^{(0)}_{l}z^{(0)}_{r} \\
{\cal V}^{z1}_{i, kl} & = & 2 \sum_{r} V_{ilkr} z^{(0)}_{r}  \;\;\;\;\; {\cal V}^{z2}_{i, kl}  =  \sum_{r} V_{iklr} z^{(0)}_{r} \\
{\cal V}^{\rho z1}_{ij,l} & = & 2 \sum_{kr} V_{iklr} z^{*(0)}_{k}
\rho^{(0)}_{rj} - V_{rklj} z^{*(0)}_{k} \rho^{(0)}_{ir} + \sum_{kr} \left [
V_{irkl} z^{(0)}_{k} + V_{irlk} z^{(0)}_{k} \right ] \kappa^{*(0)}_{rj} \\
{\cal V}^{\rho z2}_{ij,k} & = & 2 \sum_{lr} V_{iklr} z^{(0)}_{l}
\rho^{(0)}_{rj} - V_{rklj} z^{(0)}_{l} \rho^{(0)}_{ir} - \sum_{lr} \left [
V_{rjkl} z^{*(0)}_{l} + V_{rjlk} z^{*(0)}_{l} \right ] \kappa^{(0)}_{ir} \\
{\cal V}^{\kappa z1}_{ij,k} & = & 2 \sum_{lr} V_{ilkr} z^{*(0)}_{l}
\kappa^{(0)}_{rj} + V_{rklj} z^{*(0)}_{l} \kappa^{(0)}_{ir} + \sum_{lr}
\left [ V_{rjkl} z^{(0)}_{l} + V_{rjlk} z^{(0)}_{l} \right ] \rho^{(0)}_{ir}
\nonumber \\
& + & \sum_{lr} \left [ V_{irkl} z^{(0)}_{l} + V_{irlk} z^{(0)}_{l} \right ]
\rho^{*(0)}_{rj} + \sum_{l} \left [ V_{ijkl} z^{(0)}_{l} + V_{ijlk}
z^{(0)}_{l} \right ] \\
{\cal V}^{\kappa z2}_{ij,k} & = & 2 \sum_{lr} V_{iklr} z^{(0)}_{l}
\kappa^{(0)}_{rj} + V_{rlkj} z^{(0)}_{l} \kappa^{(0)}_{ir} \\
{\cal W}^{\rho h}_{ij, kl} & = & 2 \sum_{r} V_{iklr} \rho^{(0)}_{rj} -
V_{rklj} \rho^{(0)}_{ir}  \;\;\;\;\;
{\cal W}^{\rho \Delta}_{ij, kl}  =  \sum_{r} V_{irkl} \rho^{(0)*}_{rj} +
V_{rjkl} \rho^{(0)}_{ir} \\
{\cal W}^{\kappa h}_{ij, kl} & = & \sum_{r} V_{iklr} \kappa^{(0)}_{rj} \;\;\;\;\; {\cal W}^{\kappa h \dagger}_{ij, kl}  =  \sum_{r} V_{rklj}
\kappa^{(0)}_{ir} \\
{\cal W}^{\kappa \Delta}_{ij, kl} & = & \sum_{r} V_{irkl} \kappa^{(0)*}_{rj} \;\;\;\;\;
{\cal W}^{\kappa \Delta \dagger}_{ij, kl}  =  \sum_{r} V_{rjkl}
\kappa^{(0)}_{ir} .
\end{eqnarray}

In addition, we define
\begin{equation}
\zeta(t)  =  {\rm diag} \left (
\begin{array}{c}
E(t) \\
E^{*}(t) \\
\epsilon^{(-)}(t) \\
\epsilon^{(+)}(t) \\
\left [\epsilon^{(-)}(t) \right ]^{*} \\
\left [ \epsilon^{(+)}(t) \right ]^{*}
\end{array}
\right ) \vec{\psi}^{(0)} \\
\equiv \left (
\begin{array}{cccccc}
E(t) & 0 & 0 & 0 & 0 & 0\\
0 & E^{*}(t) & 0 & 0 & 0 & 0\\
0 & 0 & \epsilon^{(-)}(t) & 0 & 0 & 0\\
0 & 0 & 0 & \epsilon^{(+)}(t) & 0 & 0 \\
0 & 0 & 0 &  0 & \left [\epsilon^{(-)}(t) \right ]^{*} & 0 \\
0 & 0 & 0 &  0 & 0 & \left [ \epsilon^{(+)}(t) \right ]^{*}
\end{array}
\right ) \left (
\begin{array}{c}
\vec{z}^{(0)} \\
\vec{z}^{(0)*} \\
\vec{\rho}^{(0)} \\
\vec{\kappa}^{(0)} \\
\vec{\rho}^{(0)*} \\
\vec{\kappa}^{(0)*}
\end{array}
\right ).
\end{equation}
We have further defined the $n^2 \times n^2$ matrices
\begin{equation}
{\cal \epsilon}^{(\pm)}(t)_{ij, kl}  =  E_{ik}(t) \delta_{jl} \pm
E_{lj}(t) \delta_{ik} .
\end{equation}
As mentioned in the main text, ``${\rm diag} [A B C \cdots ]$'' denotes block diagonal square matrix with the component
matrices $A, B, C, \cdots$ as its diagonal blocks.

\section{Bogoliubov-de Gennes Equations for Contact interatomic interaction}

Griffin has provided a prescription for solving TIHFB in terms of the Bogoliubov-de Gennes Equations under the contact interatomic potential approximation\cite{Griffin}. In this section we show that self-consistent equations for $R^{(0)}$ [Eq. (\ref{eqmR})] is simply the Bogoliubov-de Gennes equations written in the trap basis, under the contact interatomic potential approximation, and summarize the numerical procedure used to find the solution to TIHFB.

The Bogoliubov-de Gennes equations are\cite{Griffin}:
\begin{eqnarray}
H^{sp} -\mu + 2U_{0} \left [ |\psi_{g}({\bf r})|^{2} + \tilde{n} ({\bf r})  \right ]
u_{i}({\bf r}) + U_{0}\left [ \psi_g^{2}({\bf r}) + \tilde{m}({\bf r}) \right ]  v_{i}({\bf r}) &=& E_{i} u_{i}({\bf r}) \nonumber \\
H^{sp} -\mu + 2U_{0} \left [ |\psi_{g}({\bf r})|^{2} + \tilde{n} ({\bf r})  \right ]
v_{i}({\bf r}) + U_{0}\left [ \psi_g^{*2}({\bf r}) + \tilde{m}^{*}({\bf r}) \right ] u_{i}({\bf r}) & = & - E_{i} v_{i}({\bf r}), \label{BogdeGennes}
\end{eqnarray}
where the quantities $\psi_g ({\bf r})$, $\tilde{n}({\bf r})$, and $\tilde{m}({\bf r})$ are as defined in Ref. \cite{Griffin} which are respectively $z({\bf r})$, $\rho({\bf r})$, and $\kappa({\bf r})$ of Eqs. (\ref{z_r}-\ref{kappa_r}) when written in terms of our variables $z_{i}$, $\rho_{ij}$, and $\kappa_{ij}$. $u_{i}({\bf r})$ and $v_{i}({\bf r})$ are the eigenstates to be calculated and can be shown to satisfy the orthogonality and symmetry relations
\begin{eqnarray}
\int u^{*}_{i}({\bf r})u_{k}({\bf r})  -  v^{*}_{i}({\bf r})v_{k}({\bf r})  & = & \delta_{ik} \\
\int u^{*}_{i}({\bf r})v_{k}({\bf r})  +  v^{*}_{i}({\bf r})u_{k}({\bf r})  & = & 0.
\end{eqnarray}

In matrix form, Eq. (\ref{BogdeGennes}) is
\begin{equation}
\tilde{\gamma} \left (
\begin{array}{cc}
h &  \Delta  \\
\Delta^{*} & h^{*}
\end{array}
\right ) \left (
\begin{array}{c}
u \\
v
\end{array}
\right ) = E \left (
\begin{array}{c}
u \\
v
\end{array}
\right ) \label{matrixBdG}
\end{equation}
where we have changed the basis from the position basis to the trap basis by introducing the matrix elements in terms of the trap eigenstates $\phi_{i}({\bf r})$ as follows:
\begin{eqnarray}
h_{ij} & = & \int \phi^{*}_{i}({\bf r)} \left \{ H^{sp} -\mu + 2U_{0} \left [ |\psi_{g}({\bf r})|^{2} + \tilde{n} ({\bf r})  \right ] \right \} \phi_{j}({\bf r)}  d {\bf r} \\
\Delta_{ij} & = & \int \phi^{*}_{i}({\bf r)} U_{0}\left [ \psi_g^{2}({\bf r}) + \tilde{m}({\bf r}) \right ] \phi_{j}({\bf r)}  d {\bf r} \\
u_{i} & = & \int \phi^{*}_{i}({\bf r)} u_{i}({\bf r}) d {\bf r} \\
v_{i} & = & \int \phi^{*}_{i}({\bf r)} v_{i}({\bf r}) d {\bf r}
\end{eqnarray}

Using the contact interaction, and Eqs. (\ref{z_r} - \ref{kappa_r}) for $\psi_g ({\bf r})$, $\tilde{n}({\bf r})$, and $\tilde{m}({\bf r})$, it is clear that $h$ and $\Delta$ coincide with those of Eqs. (\ref{h}-\ref{Delta});
%Since the matrix of Eq. (\ref{matrixBdG}) is simply the transpose of the %matrix $H$ of Eq. (\ref{BRTIHFB}) the eigenvalues are the same for both %matrices. In addition, the symmetric structure of these matrices imply that %the associated eigenvectors are also identical.
the Bogoliubov-de Gennes equations in the trap basis therefore give the same eigenvalue problem as that of diagonalizing the matrix $\tilde{\gamma} H$ of Eq. (\ref{eqmR}).

The steps to follow in solving TIHFB are therfore\cite{Griffin}:
\begin{enumerate}
\item Solve Eq. (\ref{NLSE}) for $z_i$ assuming $\rho_{ij} = \kappa_{ij} = 0$.
\item Diagonalize $H$ of Eq. (\ref{BRTIHFB})  with the current value of $z_i$, $\rho_{ij}$ $\kappa_{ij}$ Get eigenvectors U and V.
\item Calculate new $\rho_{ij}$ and $\kappa_{ij}$ using U and V:
\begin{eqnarray}
\rho_{ij}(t) & = &  \sum_{p \neq 0} \left [ U_{pi}U^{*}_{pj} + V^{*}_{pi}V_{pj} \right ] N_p + V_{pi}^{*}V_{pj} \\
\kappa_{ij}(t) & = &  \sum_{p \neq 0} \left [ U_{pi}V^{*}_{pj} + U_{pj}V^{*}_{pi} \right ] N_p + U_{pj}V^{*}_{pi}
\end{eqnarray}
where $N_{p} = \left [ \exp(\hbar \omega_{p}/kT) - 1  \right ]^{-1}$.
\item Solve Eq. (\ref{NLSE}) for $z_i$ using the calculated values of $\rho_{ij}$ and $\kappa_{ij}$.
\item Iterate: go back to Step 2.
\item Stop the iteration when the solutions $z_{i}$, $\rho_{ij}$ and $\kappa_{ij}$ converge.
\end{enumerate}

\begin{table}[h]
\begin{center}
\begin{tabular}{|c|c|c|c|}
\hline
TIHFB ($T = 0 \hbar \omega/k_B$) & TDHFB ($T = 0 \hbar \omega/k_B$) & TIHFB ($T = 10 \hbar \omega/k_B$) & TDHFB ($T = 10 \hbar \omega/k_B$) \\
\hline
$  0 $ & $ 0  $ & $ 0  $ & $ 0  $ \\
$  0 $ & $ 0  $ & $ 0  $ & $ 0.629  $ \\
   0.3477         &    0.4555  &   0.3739        &  0.6485 \\
  0.3528        &     0.7088            &     0.6607         &    0.6668  \\
   0.7104        &    0.7136            &    0.6612          &  0.8213 \\
   0.7110        &    1.0159            &    0.8733           &  0.9282 \\
   0.8521        &    1.1046            &     0.8873    &        1.0074 \\
   1.0161        &     1.1584            &    0.8908     &        1.0114  \\
 1.0163         &     1.5040            &    0.9903    &         1.0828  \\
  1.1181        &   1.5864	         &    1.0070    &         1.5068   \\
   1.1324        &   1.7735           &      1.0071    &         1.6970  \\
   1.4644        &     1.7760        &   1.5478      &       1.7666 \\
  1.4840        &    1.8211            &    1.5520     &      1.7788 \\  
 1.7743          &   1.8832      &     1.7654     &       2.0183  \\
  1.7749         &    2.1995            &   1.7660      &    2.4251 \\
 1.8281           &   2.2553            &    1.8741    &     2.4716  \\
1.8434        &    2.4843            &     1.8824      &      2.6083  \\
 1.9372         &  2.4935            &     1.9575       &     2.6454   \\
  2.1755        &   2.6050            &    2.4265       &     2.7705 \\ 
 2.1943        &     2.7907            &    2.4272       &      2.7927  \\
   2.4851        &  2.7914            &    2.5353       &      2.8689  \\
\end{tabular}
\end{center}
\caption{The lowest positive eigenvalues of ${\cal L}_{0}$ and ${\cal L}$ that correspond to the TIHFB and TDHFB respectively at zero and finite temperatures. The eigenvalues are given in units of trap energy, $\hbar \omega_{\rm trap}$}. \label{eigenlist}
\end{table}

\begin{figure}[t]
%\centerline{\psfig{height=9cm,file=newlinrespfig1.ps}}
\caption{{\protect\footnotesize $K^{(1)}(t-t_1, {\bf r}, {\bf r}_1)$ for zero temperature condensate at times $t - t' = 0/\omega_{\rm trap}, 7.2/\omega_{\rm trap}, 15.7/\omega_{\rm trap}$ given in different columns.   The top, middle, and bottom rows give the response function for the condensate, non-condensate density, and non-condensate correlation as indicated. The dashed circle  represents the spatial extent of the trapped BEC. The positions $x$ and $x'$ are given in harmonic oscillator length units. }} \label{K1rrtT0}
\end{figure}

\noindent
\begin{figure}[t]
%\centerline{\psfig{height=9cm,file=newlinrespfig2.ps}}
\caption{{\protect\footnotesize Same as in Fig. \ref{K1rrtT0}, but at finite temperature $10 \hbar\omega/k_B$.}} \label{K1rrt}
\end{figure}

\begin{figure}[t]
%\centerline{\psfig{height=9cm,file=newlinrespfig3.ps}}
\caption{{\protect\footnotesize The left column shows $K^{(1)}(t-t_1, {\bf r})$ i.e. linear response in time at zero temperature  integrated over ${\bf r}_1$ and plotted as a function of $t-t_1$ and ${\bf r}$ for zero temperature. The details for the time of evolution $t - t' = 0\pi/\omega_{\rm trap}$ to $5\pi/\omega_{\rm trap}$ are shown.  The right column shows linear response integrated over both ${\bf r}$ and ${\bf r}_1$,   $K^{(1)}(t-t_1)$, plotted as a function of $t-t_1$ from $0$ to $5\pi/\omega_{\rm trap}$; the solid, dashed and dotted lines represent the absolute value, real part and imaginary part respectively of the integrated response function $K^{(1)}(t-t_1)$.  The position $x$ is given in harmonic oscillator length units. }} \label{K1rtT0}
\end{figure}

\begin{figure}[t]
%\centerline{\psfig{height=9cm,file=newlinrespfig4.ps}}
\caption{{\protect\footnotesize  Same as in Fig. \ref{K1rtT0}, but at finite temperature  $10 \hbar\omega/k_B$.}} \label{K1rt}
\end{figure}

\begin{figure}[t]
%\centerline{\psfig{height=9cm,file=newlinrespfig5.ps}}
\caption{{\protect\footnotesize Logarithm of linear response functions in frequency integrated over both ${\bf r}$ and ${\bf r}_1$ vs. frequency. Top three panels --  zero temperature, bottom three panels -- finite temperature $10 \hbar\omega/k_B$. The frequency $\Omega_1$ is given in units of trap frequency.}} \label{K1spec}
\end{figure}

\begin{figure}[t]
%\centerline{\psfig{height=9cm,file=newlinrespfig6.ps}}
\caption{{\protect\footnotesize $\chi^{(1)}(-\Omega,\Omega, {\bf r}, {\bf r}_1)$  at zero temperature for frequencies $\Omega = 0\omega_{\rm trap}, 0.25\omega_{\rm trap}, 0.46\omega_{\rm trap}$ given in different columns.  These frequencies represent the resonant frequency corresponding to the strongest peak for the condensate at $\Omega = 0\omega_{\rm trap}$; an off-resonant frequency $\Omega = 0.25\omega_{\rm trap}$; and the resonant frequency corresponding to the second highest peak for the condensate at $\Omega = 0.46\omega_{\rm trap}$. The positions $x$ and $x'$ are given in harmonic oscillator length units. }} \label{K1rrwT0}
\end{figure}

\begin{figure}[t]
%\centerline{\psfig{height=9cm,file=newlinrespfig7.ps}}
\caption{{\protect\footnotesize $\chi^{(1)}(-\Omega,\Omega, {\bf r}, {\bf r}_1)$ at finite temperature  for frequencies $\Omega = 0.63\omega_{\rm trap}, 1.25\omega_{\rm trap}, 1.7\omega_{\rm trap}$ given in different columns.  These frequencies represent the resonant frequency corresponding to the strongest peak for the condensate at $\Omega = 0.63\omega_{\rm trap}$; an off-resonant frequency $\Omega = 1.25\omega_{\rm trap}$; and the resonant frequency corresponding to the second highest peak for the condensate at $\Omega = 1.7\omega_{\rm trap}$. The positions $x$ and $x'$ are given in harmonic oscillator length units.}} \label{K1rrw}
\end{figure}

\begin{figure}[t]
%\centerline{\psfig{height=9cm,file=newlinrespfig8.ps}}
\caption{{\protect\footnotesize  The linear  susceptibility at zero temperature integrated over ${\bf r}_1$ and plotted as a function of $\Omega_{1}$. The left column shows unscaled spectrum as a function of position; not all resonances are shown due to scaling; only the most dominant ones are represented. In the right hand column, the function has been normalized so that all the resonances have height of one. The position  $x$ is given in harmonic oscillator length units, while the frequency $\Omega_1$ are given in units of the trap frequency.}} \label{K1rwT0}
\end{figure}

\begin{figure}[t]
%\centerline{\psfig{height=9cm,file=newlinrespfig9.ps}}
\caption{{\protect\footnotesize   Same as in Fig. \ref{K1rwT0}, but at finite temperature  $10 \hbar\omega/k_B$.}} \label{K1rw}
\end{figure}

\end{document}